\documentclass[fleqn,10pt]{wlscirep}
\usepackage[utf8]{inputenc}
\usepackage[T1]{fontenc}

\usepackage{lineno}
\usepackage{here}
\usepackage{booktabs}
\usepackage{multirow}

\usepackage{xr}
\makeatletter
\newcommand*{\addFileDependency}[1]{
  \typeout{(#1)}
  \@addtofilelist{#1}
  \IfFileExists{#1}{}{\typeout{No file #1.}}
}
\makeatother

\newcommand*{\myexternaldocument}[1]{%
    \externaldocument{#1}%
    \addFileDependency{#1.tex}%
    \addFileDependency{#1.aux}%
}
\myexternaldocument{supplement}

\title{Reputation structure in indirect reciprocity under noisy and private assessment}

\author[1,2,*]{Yuma Fujimoto}
\author[1]{Hisashi Ohtsuki}
\affil[1]{Department of Evolutionary Studies of Biosystems, SOKENDAI. Shonan Village, Hayama, Kanagawa 240-0193, Japan.}
\affil[2]{Universal Biology Institute (UBI), the University of Tokyo. 7-3-1 Hongo, Bunkyo-ku, Tokyo 113-0033, Japan.}

\affil[*]{Corresponding Author: fujimoto\_yuma@soken.ac.jp}

\begin{abstract}
Evaluation relationships are pivotal for maintaining a cooperative society. A formation of the evaluation relationships has been discussed in terms of indirect reciprocity, by modeling dynamics of good or bad reputations among individuals. Recently, a situation that individuals independently evaluate others with errors (i.e., noisy and private reputation) is considered, where the reputation structure (from what proportion of individuals in the population each receives good reputations, defined as goodness here) becomes complex, and thus has been studied mainly with numerical simulations. The present study gives a theoretical analysis of such complex reputation structure. We formulate the time change of goodness of individuals caused by updates of reputations among individuals. By considering a large population, we derive dynamics of the frequency distribution of goodnesses. An equilibrium state of the dynamics is approximated by a summation of Gaussian functions. We demonstrate that the theoretical solution well fits the numerical calculation. From the theoretical solution, we obtain a new interpretation of the complex reputation structure. This study provides a novel mathematical basis for cutting-edge studies on indirect reciprocity.
\end{abstract}
\begin{document}

\flushbottom
\maketitle
%
%
\thispagestyle{empty}

\section*{Introduction}
Indirect reciprocity refers to a mechanism of evolution of giving behavior wherein a cooperator is given help not from its beneficiary but from a third party \cite{alexander1987biology,nowak1998evolution,nowak2005evolution}. Social information about others, such as reputation or gossip, plays a central role there in order to distinguish between helpful and non-helpful individuals. In our society, it is common for individuals to give good or bad reputations to each other according to how they behaved in previous social encounters. In particular, when we establish a large-scale society, in which they contact not only their relatives but inevitably many strangers, knowing reputations of such strangers is essential. For example, it has been suggested that two thirds of our conversation is about social topics \cite{emler1994gossip,dunbar1998grooming,dunbar2004gossip}, implying the importance of reputation in our daily life. Consequently, complex structure of mutual evaluation among individuals can emerge in a society, where
a variety of individuals exist. For example, some may receive good reputations from many individuals, and others may receive bad reputations from many individuals. There may also be intermediate ones who receive good and bad reputations to some equal extent. 

Reputation in indirect reciprocity is moral assessment of individuals, namely who is good and who is bad, in a world of binary reputations. Many theoretical models of indirect reciprocity have considered a situation where all in the population give the same reputation to a given individual \cite{nowak1998dynamics,leimar2001evolution,panchanathan2003tale,ohtsuki2004should,ohtsuki2006leading,ohtsuki2007global,brandt2005indirect,pacheco2006stern,suzuki2007evolution}. One of the reasons for this treatment is because the model becomes analytically tractable. Such reputation is called ``public reputation''. Under a public reputation model, we need to know how a given individual is evaluated but not by whom, which considerably simplifies the system. The reputation state of all individuals in the population is given by a one-dimensional array, each component of which is how individual (say, $i$) is evaluated. 

To consider a more realistic and general situation, however, we suppose another setting in which each individual independently evaluates a given individual. A reputation given under such a system is called ``private reputation''. Under the assumption of private reputation, opinions on the same person may not agree between individuals, and hence we need to know not only how a given one is evaluated but also by whom. The reputation state of all individuals in the population is, thus, generally represented by a two-dimensional matrix (called ``image matrix'' \cite{uchida2010effect,sigmund2012moral,uchida2013effect,okada2017tolerant,hilbe2018indirect}; each of its components represents how individual (say, $i$) is evaluated by another individual (say, $j$). 

Reputations can be private if only a part of the individuals in the population can observe a specific interaction \cite{nowak1998dynamics,brandt2004logic,uchida2010effect,uchida2013effect,okada2017tolerant,hilbe2018indirect}, if there is a possibility of individually committing errors in assigning reputations to others \cite{leimar2001evolution,brandt2004logic,uchida2013effect,yamamoto2017norm,hilbe2018indirect}, and/or if different individuals adopt different rules of reputation assessment \cite{brandt2004logic,uchida2010effect,uchida2010competition,uchida2018theoretical}. In recent years, models of private reputation have been used in studies for reasoning a variety of human nature, such as empathetic behaviors \cite{radzvilavicius2019evolution,krellner2020putting,krellner2021pleasing}, prejudicial attitudes \cite{whitaker2018indirect}, and so on \cite{brush2018indirect,lee2021local,schmid2021evolution,murase2022social}. Most of those studies have been based on individual-based computer simulations so far (but see references \cite{uchida2010competition,uchida2013effect,okada2018solution,lee2021local,lee2022second}), primarily because of its difficulty of their analytical treatment. In that respect, the study \cite{okada2018solution} is notable in the sense that it makes a strict, but extreme, assumption that a social interaction is observed by a single observer, in order for the authors to avoid solving infinitely many equations of joint probabilities. However, for a more general setting where many observers independently observe the same social interaction, the nature of ``image matrix'', that is, the opinion distribution of who evaluates whom and how, has been studied only through computer simulations. To our knowledge, no analytical insights have been provided so far.

In this study, we analytically tackle the question of private reputation. We assume that all individuals adopt the same ``discriminator strategy'' (explained in Model section in detail). Following a widespread convention in studies of indirect reciprocity, we consider a world of binary reputation; an individual is deemed either good or bad. A rule of how to assign a reputation is called ``social norm", and we assume that all individuals in the population share the same social norm. However, as a source of disagreement between different individuals on the reputation of the same target, we consider individual errors in reputation assignment; that is, each individual can independently commit an error in assigning a reputation to others. Thus, the same person (say, $i$) can be deemed good by some individuals and deemed bad by the other individuals in the population at the same time. The ``goodness" of individual $i$ is then defined as the proportion of those who regard $i$ as good among all in the population. Under this setting, we derive an integro-differential equation that describes how the frequency distribution of goodnesses in the population changes over time, and calculate its equilibrium distribution. When the population is sufficiently large, we demonstrate that the equilibrium distribution is approximated by a summation of Gaussian functions. Furthermore, we reveal that the equilibrium distribution of goodness very much differs between social norms adopted in the population. We then give intuitive interpretations to each equilibrium. We believe that this study provides a fundamental advance in the study of indirect reciprocity. In addition, the results of this study can lead to unraveling complex relationships among individuals through reputations in a society.

\section*{Model} 
Let us consider a population where there are a certain number, $N$, of individuals. Suppose that at each time $t$, either a good or bad reputation is given from every individual to every individual. This corresponds to a case of ``private reputation'', where each individual independently assigns a reputation toward the same target. Let $\beta_{ji}$ be one (resp. zero) if individual $i$ is good (resp. bad) in the eyes of individual $j$. Matrix $\{\beta_{ji}\}$ is called the image matrix. As noted in the introduction, the ``goodness'' of individual $i$, denoted by $p_{i}$, is defined as the proportion of individuals who give a good reputation to individual $i$ in the population. Thus, it is given as $p_i=N_i/N$, where $N_{i}=\sum_{j=1}^{N} \beta_{ji}$ is the total number of individuals who give a good reputation to individual $i$.

\begin{figure}[htbp]
\begin{center}
\includegraphics[width=0.35\linewidth]{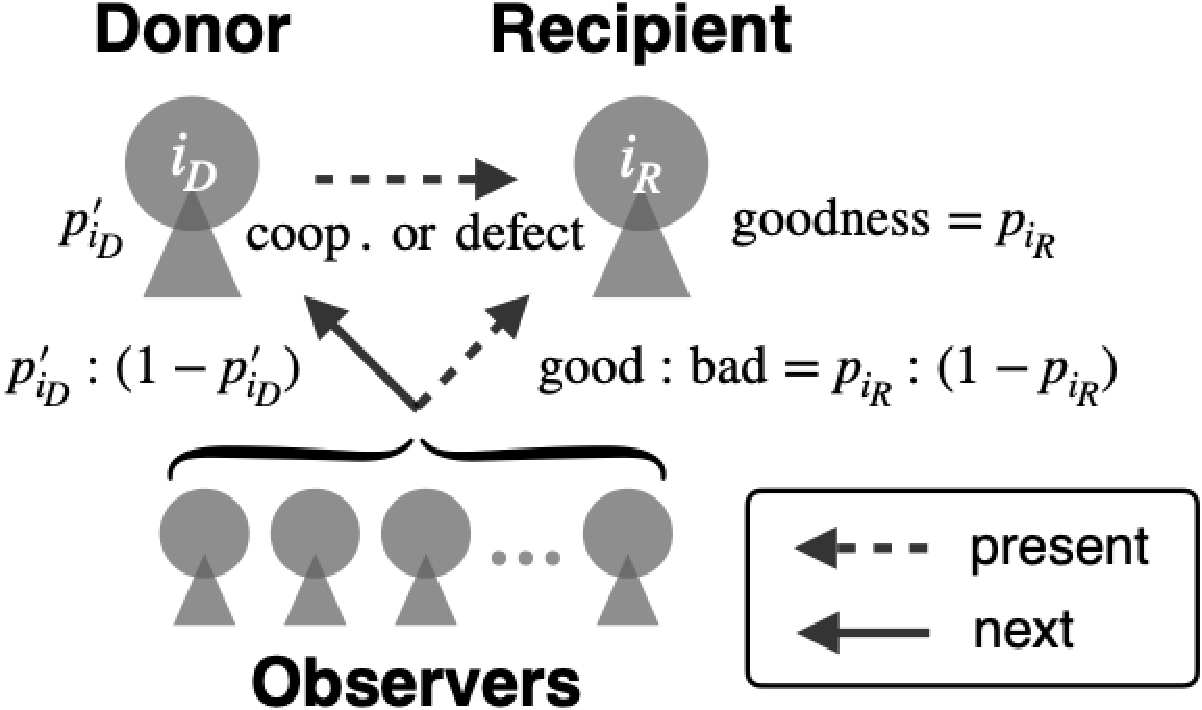}
\caption{Schematics of indirect reciprocity with private reputation. In every round, a donor ($i_D$) and a recipient ($i_R$) are randomly chosen. A goodness of the recipient in the present round is given by $p_{i_R}$. In other words, the recipient's reputation in the eyes of a random observer is good (resp. bad) with probability $p_{i_R}$ (resp. $1-p_{i_R}$). The donor chooses cooperation (resp. defection) with the recipient if the recipient's reputation in the eyes of the donor is good (resp. bad). After the interaction, each observer independently assigns a new reputation to the donor by taking into account whether the donor took cooperation or defection and whether the recipient's reputation in the eyes of that observer was good or bad before the interaction. As a result, the goodness of the donor is updated to $p'_{i_D}$.}
\label{F00}
\end{center}
\end{figure}

At each elementary step of update, we randomly select a donor and a recipient from among $N$ individuals (see Fig.~\ref{F00} for schematics). They may be the same individual, but such a case occurs with probability $1/N$ and can rightfully be neglected in the following analysis that assumes a large $N$. The donor takes one of the two actions to the recipient; cooperation or defection. The donor has a rule to choose to cooperate or defect, that can be conditional on the reputation of the recipient in the eyes of the donor. This study supposes that all individuals have the same rule called ``discriminator strategy''. A donor with this strategy chooses to cooperate (resp. defect) with a recipient whose reputation is good (resp. bad) in the eyes of the donor. In words of the image matrix, donor ($i_D$) chooses cooperation (resp. defection) with recipient ($i_R$) if $\beta_{{i_D}{i_R}}=1$ (resp. $\beta_{{i_D}{i_R}}=0$). We suppose that with probability $0 \le e_1 \le 1/2$, the donor takes the opposite action to the intended one, in which case we say that an ``error in action'' occurred.

After an action, intended or unintended, is taken by the donor, all individuals independently update the donor's reputations as observers (see Fig.~\ref{F00}). In words of the image matrix, a reputation of the donor ($i_D$) in the eyes of each observer $j$, which is denoted by $\beta_{j{i_D}}$, is updated for all $j$ simultaneously. How each observer updates donor's reputation follows a social norm adopted by the observer. In this study, we consider ``second-order'' social norms, which are mappings that assign an updated reputation to the donor from a combination of the donor's actual action (first-order information) and the recipient's reputation in the eyes of the observer (second-order information) \cite{santos2021complexity}. We assume that all individuals adopt the same social norm. Here, we suppose that with probability $0 < e_2 < 1/2$, an observer assigns to the donor the opposite reputation to the intended one, in which case we say that an ``error in assessment" occurred.

Furthermore, among $2^4=16$ possible second-order social norms, we focus on four norms: Stern-Judging (SJ), Simple Standing (SS), Shunning (SH), and Scoring (SC), which have often been the main target of studies in the literature of indirect reciprocity among 16 possible second-order norms \cite{nowak1998dynamics,pacheco2006stern,ohtsuki2007global}. Table~\ref{T01} shows how these four norms assign a reputation. Observers with these norms assign a good (resp. bad) reputation to a donor when the donor cooperates (resp. defects) with a good recipient from the observer's point of view. On the other hand, there are some differences in their ways of reputation assignment when the recipient is bad from the observer's point of view. First, SC \cite{nowak1998dynamics} gives the same reputation independent of whether the recipient is good or bad. Thus, SC is a first-order norm in accurate classification. Second, SJ \cite{pacheco2006stern} (also known as ``Kandori'' after Kandori \cite{kandori1992social}) conversely assigns a bad (resp. good) reputation to a donor who cooperates (resp. defects) with the bad recipient. Third, SS \cite{ohtsuki2007global} always gives a good reputation to a donor when the recipient is bad. Fourth, SH \cite{nowak2005evolution} always gives a bad reputation to a donor when the recipient is bad.
\begin{table}[htbp]
\centering
\begin{tabular}{cc|cc|cc|cc|cc}
    & Social norm: & \multicolumn{2}{c|}{SJ} & \multicolumn{2}{c|}{SS} & \multicolumn{2}{c|}{SH} & \multicolumn{2}{c}{SC} \\
    & ${\rm Ob.}\to{\rm Re.}$: & G & B & G & B & G & B & G & B \\
    \hline
    \multirow{2}{*}{${\rm Do.}\to{\rm Re.}$:} & C & G & B & G & G & G & B & G & G \\
    & D & B & G & B & G & B & B & B & B \\
\end{tabular}
\caption{How observers with four social norms, ${\rm SJ}$, ${\rm SS}$, ${\rm SH}$, and ${\rm SC}$, assigns a reputation to a donor when an error in assessment does not occur. Rows indicate whether the donor takes cooperation (C) or defection (D) with a recipient. Columns indicate whether the recipient's reputation is good (G) or bad (B) in the eyes of the observer.}
\label{T01}
\end{table}

We are interested in what type of structure of reputation assessment between individuals emerges in the population, and why. To this end, we will analytically derive the equilibrium distribution of ``goodness'' of individuals in the population. 

\section*{An overview of simulation results} 
\begin{figure}[htbp]
\begin{center}
\includegraphics[width=0.6\linewidth]{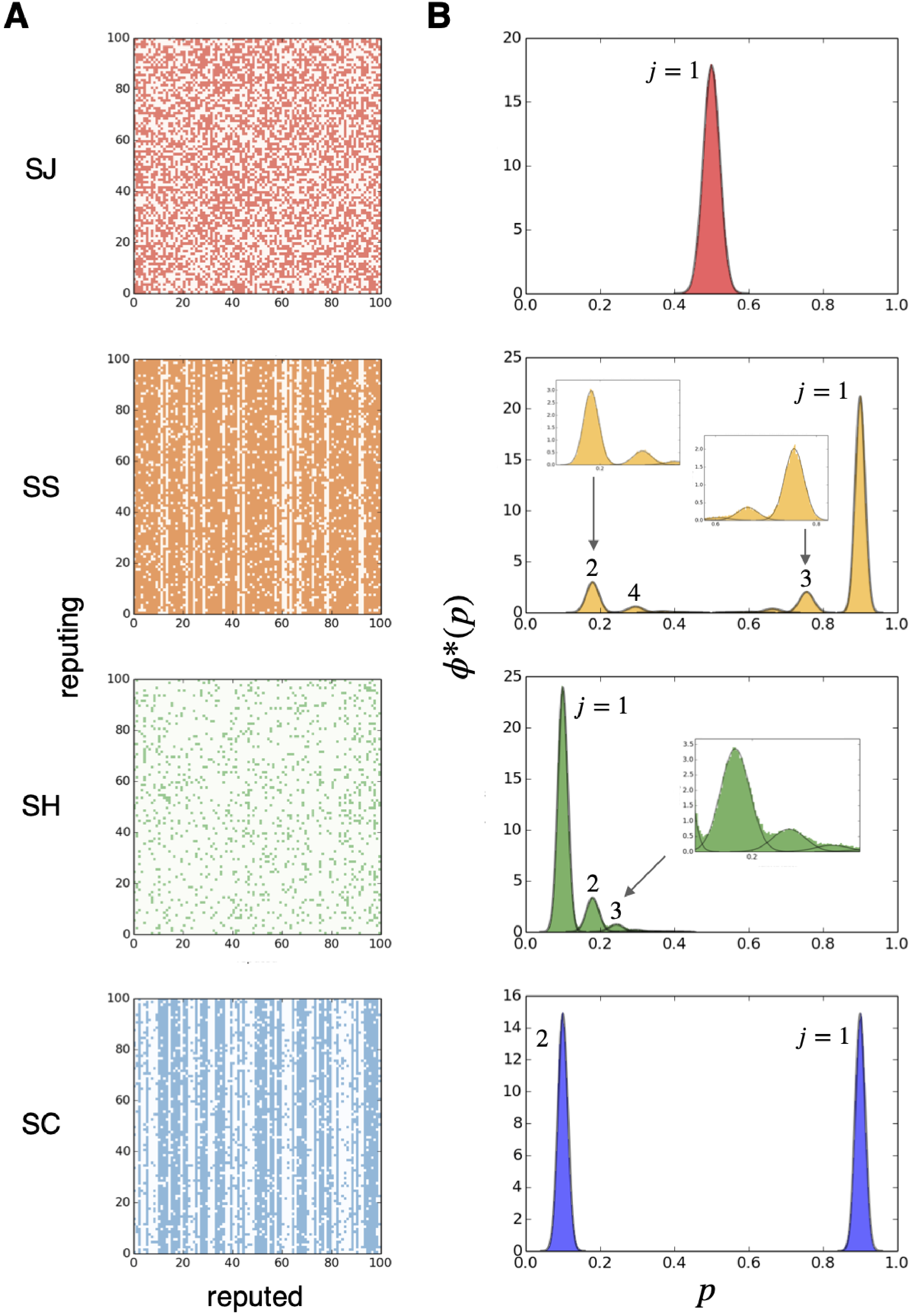}
\caption{{\bf A}. Reputations between all individuals. The image matrix $\{\beta_{ji}\}$ is drawn, where each row represents who evaluates ($j$) and each column represents who is evaluated ($i$). Colored and uncolored dots indicate good ($\beta_{ji}=1$) and bad ($\beta_{ji}=0$) reputations, respectively. From the top, each panel indicates that individuals employ norms SJ, SS, SH, and SC, respectively. One might easily see the vertical stripes on the panels of SS and SC, which mean that various goodnesses coexist among individuals. For all the panels, computer simulations are performed with parameters $N=100$, $e_1=e_2=0.1$. In our computer simulations, we assume that $N$ elementary steps of updates occur per unit time. These snapshots are taken at time $t=100$ (sufficiently long time passed). {\bf B}. Frequency distribution of goodness, $p_i$, at an equilibrium calculated from computer simulation results. The horizontal and vertical axes indicate goodness $p$ and equilibrium frequency $\phi^*(p)$, respectively. Computer simulations are performed with parameters $N=500$, $e_1=e_2=0.1$. The equilibrium frequency distribution, represented by colored areas in each panel, is calculated by taking the time average of $1000$ snapshots during time $101\le t\le 1100$. Curves in black represent our analytical approximations using mixture Gaussian distribution fitting (details explained in the main text), and they show excellent fits to the results of computer simulations (see insets for minor deviations). Numbers next to each peak represent labels of each Gaussian distribution, which shall be introduced later in the main text.}
\label{F02}
\end{center}
\end{figure}

First, we have conducted individual-based computer simulations. Fig.~\ref{F02}-A shows a snapshot of reputation assignment between all the individuals after a sufficiently long time has passed in a simulation. We note that, for social norm SJ, a similar pattern has been observed in the study \cite{uchida2013effect} (see their Fig.~2). Hilbe et al.~\cite{hilbe2018indirect} have obtained the image matrix for eight different social norms, including SS and SJ (see their Fig.~2). Fig.~\ref{F02}-B (colored area) is a frequency distribution of goodness, $p_i$, in an equilibrium state obtained by computer simulations. The four panels clearly differ from each other, depending on what social norm is employed by the population. Below we will develop a theory that explains those patterns shown in Fig.~\ref{F02}-B.

\section*{Formulation of macroscopic dynamics of reputation} 
We now consider a single update of goodness $p_i$ (see Fig.~\ref{F00} for schematics). The update is a process in which a donor and a recipient are randomly chosen from the population, the donor takes an action to the recipient, and all individuals in the population updates the donor's reputation in their eyes. Suppose that the donor is individual $i_D$ and that the recipient is individual $i_R$. In the following we denote the social norms employed in the population as $A={\rm SJ},{\rm SS},{\rm SH},{\rm SC}$.

Because the goodness of the recipient is $p_{i_R}$ and because the donor is randomly sampled from the population, the probability that the recipient is good in the eyes of the donor is $p_{i_R}$. Given this, there are two possibilities in donor's actual action toward the recipient. 

In the first possibility, the donor cooperates with the recipient. This occurs with probability
\begin{linenomath}
\begin{align}\label{donor_cooperates}
    &h(p_{i_R}):=p_{i_R}(1-e_1)+(1-p_{i_R})e_1.
\end{align}
\end{linenomath}
Here, the first term of Eq.~(\ref{donor_cooperates}) represents the case in which the recipient's reputation in the eyes of the donor is good (with probability $p_{i_R}$) and the donor succeeds in performing cooperation as intended (with probability $1-e_1$). On the other hand, the second term represents the other case in which the recipient's reputation in the eyes of the donor is bad (with probability $1-p_{i_R}$) but the donor erroneously cooperates (with probability $e_1$). When the donor cooperates with the recipient, the number of those who assign a good reputation to the donor at the next time step (i.e. after this donor's cooperation), denoted as $N_{i_D}'$, is given by
\begin{linenomath}
\begin{equation}
\begin{split}
    &N_{i_D}' = X_{1} + X_{2}, \\
    &X_{1} \sim {\cal B}(N_{i_R},a^{\rm GC}_A), \\
    &X_{2} \sim {\cal B}(N-N_{i_R},a^{\rm BC}_A),
\end{split}
\end{equation}
or, in an equivalent shorthand notation; 
\begin{align}
    &N_{i_D}'\sim
        {\cal B}(N_{i_R},a^{\rm GC}_A)+{\cal B}(N-N_{i_R},a^{\rm BC}_A).
    \label{NiD'_when_C}
\end{align}
\end{linenomath}
Here, ${\cal B}(n,p)$ represents a binomial distribution with success probability $p$ and trial number $n$. The first term on the right side of Eq.~(\ref{NiD'_when_C}) is the number of individuals who assign good reputations to donor $i_D$ at the next time step among $N_{i_R}$ observers who assign good reputations to recipient $i_R$ at the present time step. There, $a^{\rm GC}_A$ indicates the probability that an observer who assigns a good (G) reputation to the recipient at the present time step assigns a good reputation at the next time step to the donor who cooperates (C) with that recipient under social norm $A$. The values of $a^{\rm GC}_A$ can be calculated for each social norm $A$, and they are shown in Table~\ref{T02}. The second term in the right side of Eq.~(\ref{NiD'_when_C}) is the number of individuals who assign good reputations to donor $i_D$ at the next time step among $N-N_{i_R}$ observers who assign bad reputations to the recipient at the present time step. There, $a^{\rm BC}_A$ indicates the probability that an observer who assigns a bad (B) reputation to the recipient at the present time step assigns a good reputation at the next time step to the donor who cooperates (C) with that recipient under social norm $A$ (see Table~\ref{T02}). For the calculation of $a^{\rm GC}_A$ and $a^{\rm BC}_A$, compare Tables~\ref{T01} with \ref{T02}; G-pivots in Table~\ref{T01} become $1-e_{2}$ in Table~\ref{T02}, corresponding to the fact that the assignment of a good reputation to the donor is successful without an error in assessment with probability $1-e_{2}$, and B-pivots in Table~\ref{T01} become $e_{2}$ in Table~\ref{T02}, corresponding to the fact that a good reputation is erroneously assigned to the donor with probability $e_{2}$. 

\begin{table}[h!]
\centering
\begin{tabular}{cc|cc|cc|cc|cc|cc}
    & Social norm: & \multicolumn{2}{c|}{$A$} & \multicolumn{2}{c|}{SJ} & \multicolumn{2}{c|}{SS} & \multicolumn{2}{c|}{SH} & \multicolumn{2}{c}{SC} \\
    & ${\rm Ob.}\to{\rm Re.}$: & G & B & G & B & G & B & G & B & G & B \\
    \hline
    \multirow{2}{*}{${\rm Do.}\to{\rm Re.}$:} & C & $a^{\rm GC}_{A}$ & $a^{\rm BC}_{A}$ & $1-e_2$ & $e_2$ & $1-e_2$ & $1-e_2$ & $1-e_2$ & $e_2$ & $1-e_2$ & $1-e_2$ \\
    & D & $a^{\rm GD}_{A}$ & $a^{\rm BD}_{A}$ & $e_2$ & $1-e_2$ & $e_2$ & $1-e_2$ & $e_2$ & $e_2$ & $e_2$ & $e_2$ \\
\end{tabular}
\caption{Probabilities with which an observer assigns a good reputation to a donor, given the donor's action toward the recipient and the observer's evaluation of the recipient at the present time. Rows indicate whether the donor chooses to cooperate (C) or defect (D) with the recipient, and columns indicate whether the observer assigns a good (G) or bad (B) reputation to the recipient at the present time step.}
\label{T02}
\end{table}

The expected value and the variance of $p_{i_D}'=N_{i_D}'/N$ are given by
\begin{linenomath}
\begin{align}
        \textrm{E}[p_{i_D}']&=\frac{\textrm{E}[N_{i_D}']}{N}=(\underbrace{a^{\rm GC}_A-a^{\rm BC}_A}_{=:\Delta f_A^{\rm C}})p_{i_R}+a^{\rm BC}_A & &(=:f_A^{\rm C}(p_{i_R})),
        \label{f_C}\\
        \textrm{Var}[p_{i_D}']&=\frac{\textrm{Var}[N_{i_D}']}{N^2}=\frac{p_{i_R}a^{\rm GC}_A(1-a^{\rm GC}_A)+(1-p_{i_R})a^{\rm BC}_A(1-a^{\rm BC}_A)}{N}=\frac{e_2(1-e_2)}{N} & &(=:s^2) \label{s^2_C}.
\end{align}
\end{linenomath}

In the second possibility, the donor defects with the recipient. This occurs with the complementary probability to Eq.~(\ref{donor_cooperates}), that is
\begin{linenomath}
\begin{align}\label{donor_defects}
    &1-h(p_{i_R})=p_{i_R}e_1+(1-p_{i_R})(1-e_1).
\end{align}
The value of $N_{i_D}'$ at the next time step follows
\begin{align}
    &N_{i_D}'\sim
        {\cal B}(N_{i_R},a^{\rm GD}_A)+{\cal B}(N-N_{i_R},a^{\rm BD}_A),
    \label{NiD'_when_D}
\end{align}
\end{linenomath}
where we have used the same shorthand notation as Eq.~(\ref{NiD'_when_C}). Here, $a^{\rm GD}_A$ indicates the probability that an observer who assigns a good (G) reputation to the recipient at the present time step assigns a good reputation at the next time step to the donor who defects (D) with that recipient under social norm $A$. Similarly, $a^{\rm BD}_A$ indicates the probability that an observer who assigns a bad (B) reputation to the recipient at the present time step assigns a good reputation at the next time step to the donor who defects (D) with that recipient under social norm $A$. See Table~\ref{T02} for their values.  

The expected value and the variance of $p_{i_D}'=N_{i_D}'/N$ are given by
\begin{linenomath}
\begin{align}
        \textrm{E}[p_{i_D}']&=\frac{\textrm{E}[N_{i_D}']}{N}=(\underbrace{a^{\rm GD}_A-a^{\rm BD}_A}_{=:\Delta f_A^{\rm D}}) p_{i_R} +a^{\rm BD}_A & &(=:f_A^{\rm D}(p_{i_R})),
        \label{f_D}\\
        \textrm{Var}[p_{i_D}']&=\frac{\textrm{Var}[N_{i_D}']}{N^2}=\frac{p_{i_R}a^{\rm GD}_A(1-a^{\rm GD}_A)+(1-p_{i_R})a^{\rm BD}_A(1-a^{\rm BD}_A)}{N}=\frac{e_2(1-e_2)}{N} & &(= s^2) \label{s^2_D}.
\end{align}
\end{linenomath}

Two linear functions $f_A^{\rm C}$ (defined in Eq.~(\ref{f_C})) and $f_A^{\rm D}$ (defined in Eq.~(\ref{f_D})), as well as their slopes $\Delta f_A^{\rm C}$ and $\Delta f_A^{\rm D}$, will be of particular importance in the analysis below. In the following, we call $f_A^{\rm C}$ and $f_A^{\rm D}$ ``C-map'' and ``D-map'', respectively.

\section*{Time change of reputation distribution} 
For simplicity we start with the case of $N\to\infty$, where the variance $s^2$ in Eq.~(\ref{s^2_C}) and Eq.~(\ref{s^2_D}) is ignored. Let us define $\phi(p)$ as a frequency distribution of individuals with goodness $p$ in the population. Then, its time evolution is given by
\begin{linenomath}
\begin{equation}
    \frac{\mathrm{d}}{\mathrm{d}t}\phi(p)=-\phi(p)+\int_{0}^{1}\{h(p')\delta(p-f_A^{\rm C}(p'))+(1-h(p'))\delta(p-f_A^{\rm D}(p'))\}\phi(p')\mathrm{d}p'.
    \label{phi_p}
\end{equation}
\end{linenomath}
Here, we use $\delta(\cdot)$ as a Dirac delta function. In Eq.~(\ref{phi_p}), the first term on the right side represents a loss of individuals with goodness $p$ due to updates of their reputations. The first (resp. second) term inside the integral on the right side represents donors with an updated goodness $p$ after meeting a recipient with goodness $p'$ and cooperating (resp. defecting) with him/her.

Next, we consider a case of $1\ll N<\infty$, and replace delta functions in Eq.~(\ref{phi_p}) with Gaussian functions, because binomial distribution is well approximated by Gaussian distribution for large $N$. In the following, we represent a Gaussian function with mean $\mu$ and variance $\sigma^2$ by
\begin{linenomath}
\begin{equation}
    g(p;\mu,\sigma^2):=\frac{1}{\sqrt{2\pi\sigma^2}}\exp\left[\frac{(p-\mu)^{2}}{2\sigma^2}\right].
\end{equation}
Accordingly, $\delta(p-f_A^{\rm C}(p'))$ and $\delta(p-f_A^{\rm D}(p'))$ in Eq.~(\ref{phi_p}) are replaced with $g(p;f_A^{\rm C}(p'),s^2)$ and $g(p;f_A^{\rm D}(p'),s^2)$, respectively. Thus we obtain
\begin{equation}
    \frac{\mathrm{d}}{\mathrm{d}t}\phi(p)=-\phi(p)+\int_{0}^{1}\{h(p')g(p;f_A^{\rm C}(p'),s^2)+(1-h(p'))g(p;f_A^{\rm D}(p'),s^2)\}\phi(p')\mathrm{d}p'.
    \label{phi_p2}
\end{equation}
\end{linenomath}

\section*{A calculation of equilibrium state} 
Again, we start with the case of $N\to\infty$. When ${\rm d}\phi/{\rm d}t=0$ is satisfied in Eq.~(\ref{phi_p}), an equilibrium state $\phi=\phi^*$ of the equation is given by
\begin{linenomath}
\begin{equation}
    \phi^*(p)=\int_{0}^{1}\{h(p')\delta(p-f_A^{\rm C}(p'))+(1-h(p'))\delta(p-f_A^{\rm D}(p'))\}\phi^*(p')\mathrm{d}p'.
\label{equilibrium_d}
\end{equation}
We assume that this equilibrium state is described by a summation of delta functions with peak $\mu_j$ and mass $q_j$ ($j=1, \cdots)$, i.e.,
\begin{equation}
\begin{split}
    &\phi^*(p)=\sum_jq_j\delta(p-\mu_j),\\
    &\sum_jq_j=1.
\end{split}
\label{solution_d}
\end{equation}
By substituting Eq.~(\ref{solution_d}) in Eq.~(\ref{equilibrium_d}), we obtain
\begin{equation}
\begin{split}
    \sum_jq_j\delta(p-\mu_j)&=\int_{0}^{1}\{h(p')\delta(p-f_A^{\rm C}(p'))+(1-h(p'))\delta(p-f_A^{\rm D}(p'))\}\sum_jq_j\delta(p'-\mu_j)\mathrm{d}p'\\
    &=\sum_jq_j\{h(\mu_j)\delta(p-f_A^{\rm C}(\mu_j))+(1-h(\mu_j))\delta(p-f_A^{\rm D}(\mu_j))\}.
\end{split}
\label{equilibrium_d2}
\end{equation}
Thus, the problem in the case of $N\to\infty$ is to derive pairs $\{(q_j,\mu_j)\}_{j=1, \cdots}$ which satisfy Eq.~(\ref{equilibrium_d2}).

Next, we consider the case of $1\ll N<\infty$. From Eq.~(\ref{phi_p2}), an equilibrium state $\phi=\phi^*$ is given by
\begin{equation}
    \phi^*(p)=\int_{0}^{1}\{p'g(p;f_A^{\rm C}(p'),s^2)+(1-p')g(p;f_A^{\rm D}(p'),s^2)\}\phi^*(p')\mathrm{d}p',
\label{equilibrium_g}
\end{equation}
Here we assume that this equilibrium state is given by a summation of Gaussian functions;
\begin{equation}
\begin{split}
    &\phi^*(p)=\sum_jq_jg(p;\mu_j,\sigma_j^2),\\
    &\sum_jq_j=1.
\end{split}
\label{solution_g}
\end{equation}
We further assume that deviations $\sigma_j$ are negligible in the order of $O(1)$ and that
\begin{equation}
    \sigma_j=O(N^{-1/2})
\end{equation}
holds. When we substitute Eq.~(\ref{solution_g}) into Eq.~(\ref{equilibrium_g}), we obtain
\begin{equation}
\begin{split}
    \sum_jq_jg(p;\mu_j,\sigma_j^2)&=\int_{0}^{1}\{h(p')g(p;f_A^{\rm C}(p'),s^2)+(1-h(p'))g(p;f_A^{\rm D}(p'),s^2)\}\sum_jq_jg(p';\mu_j,\sigma_j)\mathrm{d}p'\\
    &=\sum_jq_j\int_{0}^{1}\{h(p')g(p;f_A^{\rm C}(p'),s^2)+(1-h(p'))g(p;f_A^{\rm D}(p'),s^2)\}g(p';\mu_j,\sigma_j^2)\mathrm{d}p'\\
    &\simeq\sum_jq_j\int_{-\infty}^{\infty}\{h(\mu_j)g(p;f_A^{\rm C}(p'),s^2)+(1-h(\mu_j))g(p;f_A^{\rm D}(p'),s^2)\}g(p';\mu_j,\sigma_j^2)\mathrm{d}p'\\
    &=\sum_jq_j\{h(\mu_j)g(p;f_A^{\rm C}(\mu_j),s^2+(\Delta f_A^{\rm C})^2\sigma_j^2)+(1-h(\mu_j))g(p;f_A^{\rm D}(\mu_j),s^2+(\Delta f_A^{\rm D})^2\sigma_j^2)\}.
\end{split}
\label{equilibrium_g2}
\end{equation}
\end{linenomath}
Here, from the second to third line, we have used the following two approximations. One is that the interval of integral $0\le p'\le 1$ is replaced with $-\infty < p' < \infty$. The other is that some but not all $p'$ are replaced with $\mu_j$. A rationale behind these approximations are that $g(p';\mu_j,\sigma_j^2)$ is almost zero outside the interval $\mu_j - O(N^{-1/2}) < p' < \mu_j + O(N^{-1/2})$, the width of which is as small as $O(N^{-1/2})$. From the third to fourth line in Eq.~(\ref{equilibrium_g2}), we have calculated an integral of a product of two Gaussian functions through completing the square with respect to $p'$, as
\begin{linenomath}
\begin{equation}
\begin{split}
    &\int_{-\infty}^{\infty}g(p;f_A^{\rm C}(p'),s^2)g(p';\mu_j,\sigma_j^2)\mathrm{d}p'\\
    &=\int_{-\infty}^{\infty}\frac{1}{\sqrt{2\pi s^2}}\exp\left[-\frac{(p-f_A^{\rm C}(p'))^2}{2s^2}\right]\frac{1}{\sqrt{2\pi\sigma_j^2}}\exp\left[-\frac{(p'-\mu_j)^2}{2\sigma_j^2}\right]\mathrm{d}p'\\
    &=\frac{1}{\sqrt{2\pi s^2}}\frac{1}{\sqrt{2\pi\sigma_j^2}}\int_{-\infty}^{\infty}\exp\left[-\frac{(p-(\Delta f_A^{\rm C}p' + a_A^{\rm BC}) )^2}{2s^2}-\frac{(p'-\mu_j)^2}{2\sigma_j^2}\right]\mathrm{d}p'\\
    &=\frac{1}{\sqrt{2\pi s^2}}\frac{1}{\sqrt{2\pi\sigma_j^2}}\exp\left[-\frac{(p-(\Delta f_A^{\rm C}\mu_j + a_A^{\rm BC}) )^2}{2(s^2+(\Delta f_A^{\rm C})^2\sigma_j^2)}\right]\underbrace{\int_{-\infty}^{\infty}\exp\left[-\frac{s^2+(\Delta f_A^{\rm C})^2\sigma_j^2}{2s^{2}\sigma_j^2}\left(p'-\frac{s^2\mu_j+(\Delta f_A^{\rm C})(p-a_A^{\rm BC})\sigma_j^2}{s^2+(\Delta f_A^{\rm C})^2\sigma_j^2} \right)^2\right]\mathrm{d}p'}_{=\displaystyle \sqrt{2\pi\frac{s^2\sigma_j^2}{s^2+(\Delta f_A^{\rm C})^2\sigma_j^2}}}\\
    &=\frac{1}{\sqrt{2\pi(s^2+\sigma_j^2(\Delta f_A^{\rm C})^2)}}\exp\left[-\frac{(p-f_A^{\rm C}(\mu_j))^2}{2(s^2+(\Delta f_A^{\rm C})^2\sigma_j^2)}\right]\\
    &=g(p;f_A^{\rm C}(\mu_j),s^2+(\Delta f_A^{\rm C})^2\sigma_j^2).
\end{split}
\end{equation}
\end{linenomath}
Thus, the problem in the case of $1\ll N<\infty$ is to derive triples $\{(q_j,\mu_j,\sigma_j)\}_{j=1, \cdots}$ which satisfy Eq.~(\ref{equilibrium_g2}).

Now we give an intuitive interpretation of Eq.~(\ref{equilibrium_g2}). The left side of Eq.~(\ref{equilibrium_g2}) represents a summation of Gaussian functions with mean $\mu_j$ and variance $\sigma_j^2$, whereas the right side represents another summation of Gaussian functions, which have been transformed from the original summation. Let us call individuals represented by the $j$-th Gaussian function $g(p;\mu_j,\sigma_j^2)$ with mass $q_{j}$ ``class-$j$'' individuals. Eq.~(\ref{equilibrium_g2}) tells us that among those donors who interact with class-$j$ recipients, the fraction $h(\mu_j)$ of them cooperate with their recipients, and the distribution of their updated goodness becomes $g(p;f_A^{\rm C}(\mu_j),s^2+(\Delta f_A^{\rm C})^2\sigma_j^2)$. The transition of mean, $\mu_j\mapsto f_A^{\rm C}(\mu_j)$, is governed by the C-map. As for the transition of variance, $\sigma_j^2\mapsto s^2+(\Delta f_A^{\rm C})^2\sigma_j^2$, the first term $s^2$ represents newly generated variance due to errors in assessment and to the finiteness of the population size. The second term $(\Delta f_A^{\rm C})^2\sigma_j^2$ means that the variance $\sigma_j^2$ that originally existed in the distribution of goodness of class-$j$ recipients is damped by the C-map (recall that its slope is $\Delta f_A^{\rm C}$). Similarly, among those donors who interact with class-$j$ recipients, the fraction $1-h(\mu_j)$ of them defect with their recipients, and the distribution of their updated goodness becomes $g(p;f_A^{\rm D}(\mu_j),s^2+(\Delta f_A^{\rm D})^2\sigma_j^2)$. Similar explanations are possible for the transition of mean, $\mu_j\mapsto f_A^{\rm D}(\mu_j)$ and for the transition of variance, $\sigma_j^2\mapsto s^2+(\Delta f_A^{\rm D})^2\sigma_j^2$.

\subsection*{Equilibrium state for each social norm} 
We now give an overview of our calculation of the equilibrium state for each social norm, $A={\rm SJ},{\rm SS},{\rm SH},{\rm SC}$. Fig.~\ref{F02}-B shows that analytical solutions to Eq.~(\ref{equilibrium_g2}) excellently fit results of computer simulations (see SI for a more detailed calculation of $(q_j, \mu_j,\sigma_j)$).

\begin{figure}[htbp]
\begin{center}
\includegraphics[width=0.8\linewidth]{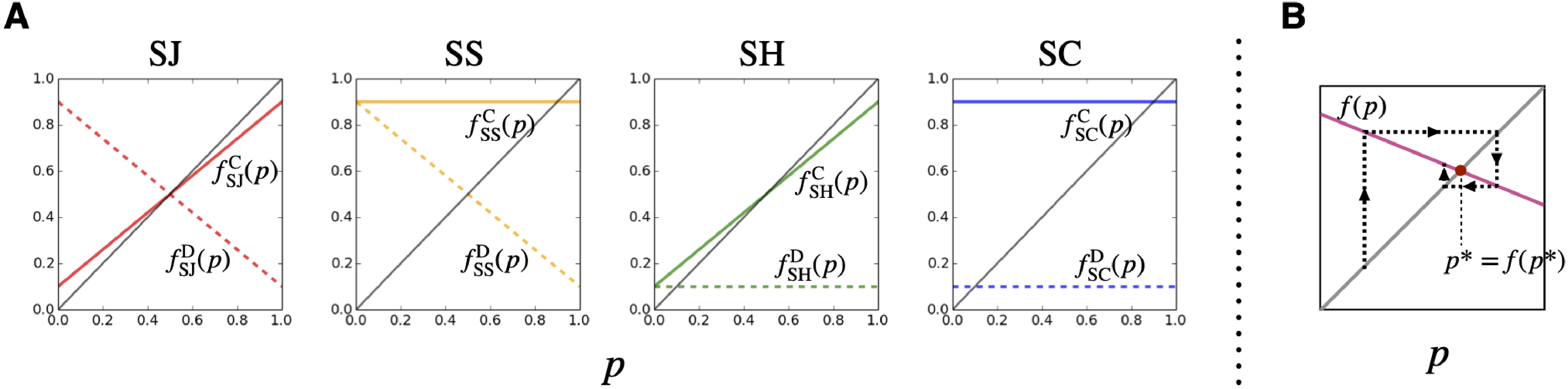}
\caption{{\bf A}. C-map $f_{A}^{\rm C}$ (solid line) and D-map $f_{A}^{\rm D}$ (broken line). We have used $e_1=e_2=0.1$. From the left to right, the panels show cases of $A={\rm SJ},{\rm SS},{\rm SH},{\rm SC}$. Black solid line indicates an identity map.  {\bf B}. Illustration of reaching a fixed point $p^*=f(p^*)$ by sequential application of a map $f$. Because slopes of all C-maps and D-maps are less than $1$ and greater than $-1$, the fixed point is always stable.}
\label{F01}
\end{center}
\end{figure}

As seen in Eq.~(\ref{equilibrium_g2}), C-map ($f_{A}^{\rm C}$) and D-map ($f_{A}^{\rm D}$) play an important role in considering the transition of each peak position $\mu_j$. As Fig.~\ref{F01}-A illustrates, these C-map and D-map differ depending on the social norm that the population adopts. If there was only one map $f$, sequentially applying this map would lead to a fixed point, which is a crossing point of map $f$ and the identity map (represented by a 45 degree line), as Fig.~\ref{F01}-B illustrates, and this fixed point would correspond to the peak position of the single Gaussian distribution at an equilibrium state. In our case, we have two maps $f_{A}^{\rm C}$ and $f_{A}^{\rm D}$ so the situation is different, but analyzing a fixed point of each map is still crucial for analyzing Eq.~(\ref{equilibrium_g2}).

Below we will study each social norm.

\begin{figure}[htbp]
\begin{center}
\includegraphics[width=0.5\linewidth]{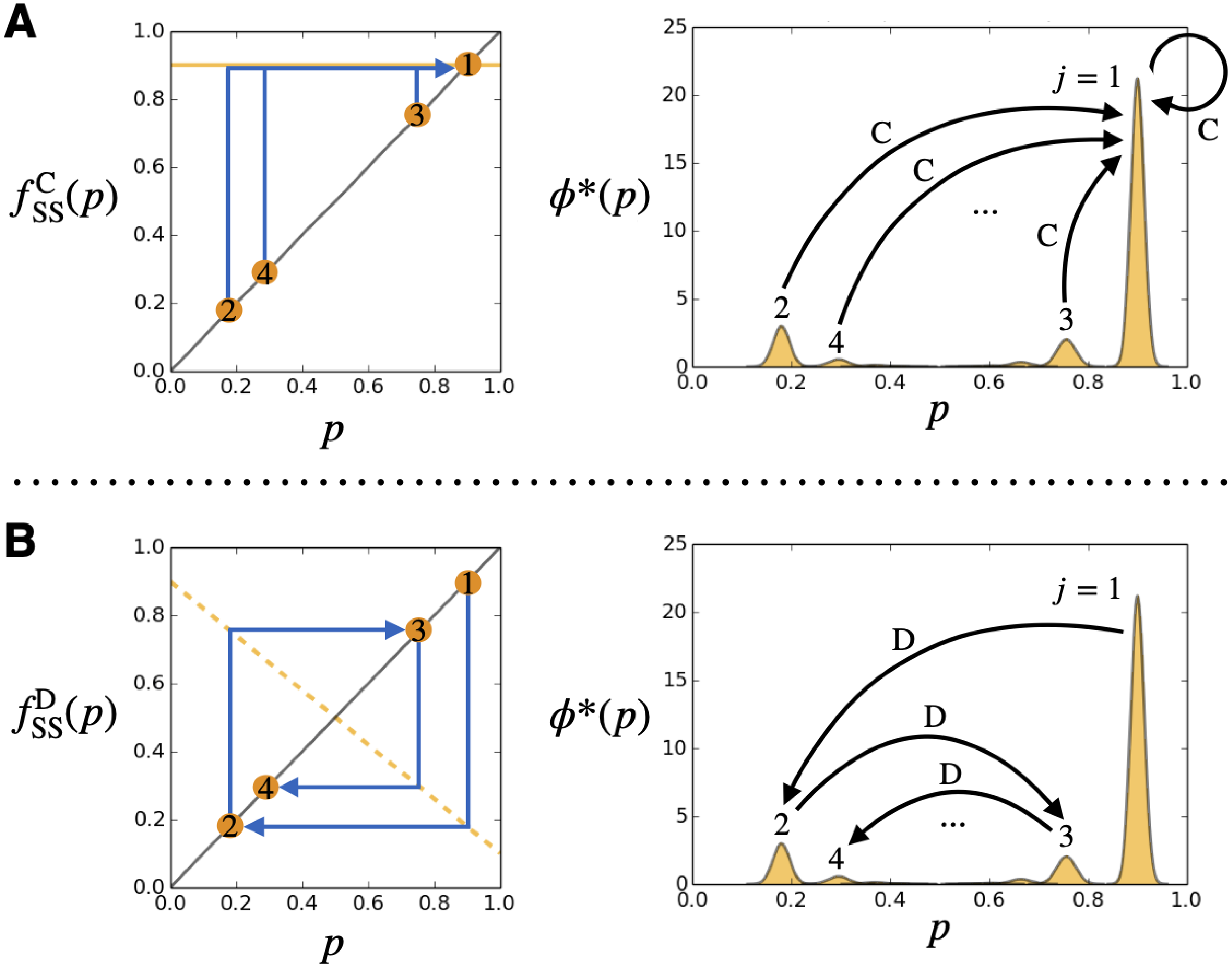}
\caption{An illustration to interpret the equilibrium state for $A={\rm SS}$. {\bf A}. In the left panel, the yellow solid line shows the C-map, which maps any value $p$ to a constant value, $1-e_2$. This mapped value is labeled as $\mu_1$. The right panel (same as a panel in Fig.~\ref{F02}-B) shows the equilibrium state $\phi^*(p)$ for $A={\rm SS}$, where the peak positions of all the classes $\mu_{1}, \mu_{2}, \mu_{3}, \cdots$ are mapped to $\mu_1$ by the C-map. {\bf B}. In the left panel, the yellow broken line shows the D-map, which sequentially maps the $1$st peak to $2$nd, $3$rd, $4$th peaks, and so on. The right panel illustrates how the peak position $\mu_j$ of class-$j$ is mapped to the peak position $\mu_{j+1}$ of class-$(j+1)$ by the D-map.}
\label{F05}
\end{center}
\end{figure}

{\bf When the social norm is SJ}: As Fig.~\ref{F01}-A shows, both C-map and D-map have the same fixed point, $p=1/2$. Thus, the only possible peak position of Gaussian distributions at the equilibrium state is at
\begin{linenomath}
\begin{equation}
    \mu_1=\frac{1}{2}.
\end{equation}
The equilibrium distribution is given by a single Gaussian distribution.

{\bf When the social norm is SS}: As Fig.~\ref{F01}-A shows, the C-map is a constant map, $f_{\rm SS}^{\rm C}(p)=1-e_2$, so this position is one of the peaks of the Gaussian distributions at the equilibrium state;
\begin{equation}
    \mu_1=1-e_2,
\end{equation}
(see an illustration in Fig.~\ref{F05}-A). The other peaks can be obtained by repeatedly applying the D-map. More specifically, $(j+1)$-th peak position $\mu_{j+1}$ is obtained by 
\begin{equation}
    \mu_{j+1}=f_{\rm SS}^{\rm D}(\mu_j)\ \ (j\ge 1),
\end{equation}
\end{linenomath}
(see an illustration in Fig.~\ref{F05}-B). These infinite classes allow us to characterize the population.

{\bf When the social norm is SH}: As Fig.~\ref{F01}-A shows, the D-map is a constant map, $f_{\rm SH}^{\rm D}(p)=e_2$, so this position is one of the peaks of the Gaussian distributions at the equilibrium state;
\begin{linenomath}
\begin{equation}
    \mu_1=e_2.
\end{equation}
The other peaks can be obtained by repeatedly applying the C-map for the same reason as in the case of $A={\rm SS}$. Thus, $(j+1)$-th peak position $\mu_{j+1}$ is obtained by
\begin{equation}
    \mu_{j+1}=f_{\rm SH}^{\rm C}(\mu_j)\ \ (j\ge 1),
\end{equation}

{\bf When the social norm is SC}: Both C-map and D-map are constant maps; $f_{\rm SC}^{\rm C}(p)=1-e_2$ and $f_{\rm SC}^{\rm D}(p)=e_2$. Thus, there are two possible peak positions of Gaussian distributions at the equilibrium state. They are at
\begin{equation}
\begin{split}
    &\mu_1=1-e_2,\\
    &\mu_2=e_2,
\end{split}
\end{equation}
\end{linenomath}
and the equilibrium distribution is given by a summation of two Gaussian distributions.

We can also derive $\sigma_j^2$ and $q_j$ for each social norm. See SI for the detailed calculation. Here we only summarize the results in Table.~\ref{TS01}.
\renewcommand{\arraystretch}{2.5}
\begin{table}[htbp]
\centering
\begin{tabular}{c|c|c|c|c}
    social norm & \# of Gaussians used & mass, $q_j$ & mean, $\mu_j$ & variance, $\sigma_j^2$  \\
    \hline
    SJ & $1$ & 1 & $\displaystyle\frac{1}{2}$ & $\displaystyle\frac{1}{4N}$  \\
    SS & $\infty$ & $\displaystyle\frac{\prod_{k=1}^{j-1}(1-h(\mu_k))}{\sum_{\ell=1}^{\infty}\prod_{k=1}^{\ell-1}(1-h(\mu_k))}$ & $\displaystyle\frac{1-\{-(1-2e_2)\}^{j}}{2}$ & $\displaystyle\frac{1-(1-2e_2)^{2j}}{4N}$ \\
    SH & $\infty$ & $\displaystyle\frac{\prod_{k=1}^{j-1}h(\mu_k)}{\sum_{\ell=1}^{\infty}\prod_{k=1}^{\ell-1}h(\mu_k)}$ & $\displaystyle\frac{1-(1-2e_2)^{j}}{2}$ & $\displaystyle\frac{1-(1-2e_2)^{2j}}{4N}$  \\
    SC & $2$ & $\displaystyle\frac{1}{2}$ & $\mu_1=1-e_2, \mu_2=e_2$ & $\displaystyle\frac{e_2(1-e_2)}{N}$  \\
\end{tabular}
\caption{Analytical solutions to Eq.~(\ref{equilibrium_g2}). $h$ is defined as $h(p)=p(1-e_1)+(1-p)e_1$ (see Eq.~(\ref{donor_cooperates})). We employ the convention, $\prod_{k=1}^{0} \cdot = 1$. From this table, we see that, for SJ norm, neither the error rate in action $e_1$ nor the error rate in assessment $e_2$ influences the stationary distribution. For SS and SH, $e_1$ influences only masses $q_j$, and $e_2$ influences masses $q_j$, means $\mu_j$, and variances $\sigma_j^2$. For SC, $e_1$ influences nothing, but $e_2$ influences means $\mu_j$ and variances $\sigma_j^2$.
}
\label{TS01}
\end{table}
\renewcommand{\arraystretch}{1.0}

Based on Table.~\ref{TS01}, we now describe the equilibrium distribution of goodness in the population for each social norm.

{\bf When the social norm is SJ}: All individuals receive good reputations from almost a half of the population and receive bad reputations from almost the other half of the population. The average goodness in the population is $1/2$. Surprisingly, the two error rates $e_1$ and $e_2$ do not affect the equilibrium distribution at all. 

{\bf When the social norm is SS}: There are an infinite number of peaks in the equilibrium distribution. The highest one is at $\mu_{1}=1-e_{2}$ and individuals that belong to this class-1 are those who receive good reputations the most. The second highest peak is at $\mu_{2} = 2e_{2}(1-e_{2})$ and individuals that belong to this class-2 are those who receive bad reputations the most. The positions of the third, fourth highest peaks and so on are arranged in an oscillating fashion across $1/2$ as $\mu_2 < \mu_4 < \cdots < 1/2 < \cdots < \mu_3 < \mu_1$. The average goodness in the population is relatively high compared with the other three social norms.

{\bf When the social norm is SH}: There are an infinite number of peaks in the equilibrium distribution. The highest one is at $\mu_{1}=e_{2}$ and individuals that belong to this class-1 are those who receive a good reputation the least. The positions of the second, third highest peaks and so on are monotonically increasing as $\mu_1 < \mu_2 < \mu_3 < \cdots < 1/2$. The average goodness in the population is relatively low compared with the other three social norms.

{\bf When the social norm is SC}: A half of the individuals receive good reputations from a majority of individuals (i.e., high goodness, $\mu_1=1-e_2$), and the other half receive bad reputations from a majority of individuals (i.e., low goodness, $\mu_2=e_2$). The average goodness in the population is $1/2$, which is the same as in the case of SJ. However, there is a large difference in the frequency distribution of goodnesses between SJ and SC, as shown in Fig.~\ref{F02}-B. The action error rate, $e_1$, does not affect the equilibrium distribution at all.

\begin{figure}[!t]
\begin{center}
\includegraphics[width=0.6\linewidth]{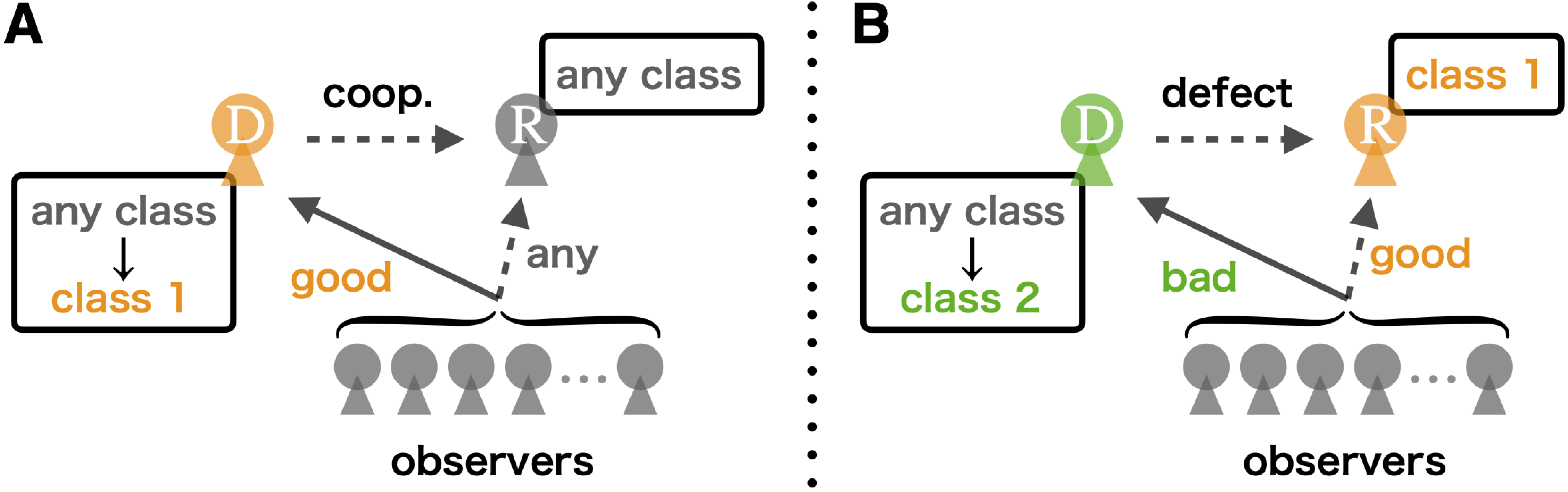}
\caption{An interpretation of the equilibrium state for SS. {\bf A}. When a donor cooperates with a recipient, the donor receives good reputations from a lot of observers, independent of classes of the donor and the recipient. Such a donor moves to class-$1$. Because this process frequently occurs, SS generates a majority of individuals with high goodness. {\bf B}. When a donor defects with a recipient in class-$1$, the donor receives bad reputations from a lot of observers and such a donor moves to class-$2$. This process does not frequently occur, but SS definitely generates a minority of individuals with low goodness.}
\label{F06}
\end{center}
\end{figure}

{\bf Remarks on SS}: The equilibrium distribution of goodness under SS is especially interesting because there are some individuals with low goodness (such as class-$2$) although the average goodness in the population is high. Here, we explain a mechanism of how such an equilibrium distribution is formed under SS. First of all, SS tends to generate many individuals with high goodnesses labeled as class-$1$. This is because once a donor cooperates with a recipient, observers assign good reputations to the donor under SS regardless of whether the recipient's reputation in the eyes of those observers is good or bad, unless observers commit an error in assessment (see Fig.~\ref{F06}-A). On the other hand, SS also generates a small number of individuals with low goodness labeled as class-$2$. Such individuals with low goodness emerge when a donor defects with a recipient in class-$1$, either because the donor belongs to a minority of individuals who think the recipient is bad or because the donor thinks the recipient is good but this donor erroneously chooses defection as opposed to his/her intention. In either case, such a donor receives bad reputations from almost all observers and descend to class-$2$, because in the eyes of those observers the donor's defection is seen as a defection against a good recipient (see Fig.~\ref{F06}-B). The mechanism of how individuals in class-$(j+1)$ are generated is similar; a donor who defects with a recipient in class-$j$ moves to class-$(j+1)$. 

One might expect that the goodness of a randomly sampled individual from the population that employs SS should always be higher than the goodness of a randomly sampled individual from the population that employs SJ, because SS assigns a good reputation in more cases than SJ (compare SS an SJ in Table~\ref{T01}; if a donor receives a good reputation under SJ, such a donor would also receive a good reputation under SS). However, this naive expectation is not true because class-$2$ individuals (and more generally, class-$2j$ individuals) under SS have the goodness of less than $1/2$, whereas all individuals under SJ have the goodness of about $1/2$. This apparently paradoxical phenomenon is now explained as follows. Observers under SS more frequently assign good reputations than those under SJ, and thus generate a lot of individuals with high goodness (i.e., class-$1$ individuals). However, the existence of such individuals in turn causes the emergence of a minority of individuals with low goodness (such as class-$2$ individuals). As a result, a large divide with respect to one's goodness occurs among individuals in the population that employs SS; extremely good individuals and extremely bad individuals coexist there.

\section*{Discussion and conclusion} 
This study theoretically analyzed a question of how reputation relationships among individuals are established, by using a model of indirect reciprocity. Under the assumption of private reputations, the question has mainly been discussed by computer simulations until now, except for few studies \cite{uchida2013effect,okada2018solution}. Here we formulated a change of ``goodness'' of an individual, which is defined as the proportion of individuals who regard the focal individual as good, by a stochastic process (Eqs.~(\ref{NiD'_when_C}) and (\ref{NiD'_when_D})). Then, we formulated time evolution of a frequency distribution of goodness in the population by a deterministic integro-differential equation (Eq.~(\ref{phi_p2})). By employing an approximation that uses a mixture Gaussian distribution and assumes a large population size, we obtained a closed equation that the equilibrium distribution must satisfy (Eq.~(\ref{equilibrium_g2})). We succeeded in calculating the equilibrium distribution of goodness (Table.~\ref{TS01}) and interpreted its meaning.

As far as we know, this is the first study that has analytically derived the equilibrium frequency distribution of goodness for a model of indirect reciprocity that assumes private reputations, and we believe that our study provides a major advance in theoretical studies of indirect reciprocity. As relevant literature, we compare our approach with two recent works that have analytically studied a model of private reputations.

Uchida and Sasaki \cite{uchida2013effect} analyzed the \textit{average} goodness in the population under the SJ social norm for a model of private assessment, and reached the conclusion that it is $1/2$. In contrast, our study has derived the \textit{distribution} of goodness. From this obtained distribution it is easy to calculate the average goodness under SJ, that is $1/2$. Moreover, we have analyzed three other social norms, SS, SH, and SC. By using the approach developed in this paper, it is possible to analyze the other 12 second-order social norms that have not been studied here in a similar manner.  

Okada et al.~\cite{okada2018solution} studied cases where there is always only one observer who updates his/her private reputation of a donor. By making such an extreme assumption, the authors successfully avoided calculating higher-order correlations between reputations of the same individual among observers. In contrast, we have assumed that all individuals in the population play a role of observers and update their private reputation of the same donor simultaneously. Such an approach explicitly considers correlations in opinions among observers. It will be interesting to develop a similar theoretical framework to ours that studies a model in which only a part of the individuals in the population (say, proportion $0 < \theta < 1$) become observers and simultaneously update their private reputations of the same donor. We leave it as a future study. 

As significant progress in the analysis of reputation structure, this study treated a model that all the players adopt the {\bf (1) same} {\bf (2) second-order} social norms under {\bf (3) random} interactions between a donor and a recipient. However, this simple model may not perfectly reflect a real human society. First, the society consists of various kinds of people who have different viewpoints, i.e., different social norms. This extension brings another question of which social norms can be evolutionarily advantageous, concerning studies on the emergence of cooperation \cite{hamilton1964genetical,axelrod1981evolution,nowak2006five,ohtsuki2006leading} and exploitation \cite{press2012iterated,fujimoto2019emergence,fujimoto2021exploitation}. Second, real people may take more information into account than second-order social norms do when they assign reputations to others, such as third-order information (i.e. the current reputation of a donor) (e.g., social norms named standing \cite{sugden1986economics}, staying \cite{sasaki2017evolution} and consistent standing \cite{hilbe2018indirect}) or more \cite{santos2018social,santos2021complexity}. Such additional pieces of information will bring more complexity to the reputation structure among people\cite{santos2018social,santos2021complexity}. Third, real people interact mainly with neighbors or friends. Such biased interactions are often modeled by introducing lattices or complex networks\cite{fu2008reputation,traag2011indirect,masuda2012ingroup,nakamura2012groupwise,oishi2013group,tian2016cooperation,gross2019rise,dong2019second,roberts2021benefits,oishi2021balanced,santos2021social,podder2021local}. Our study can be applied to the analysis of reputation structure even for such extended situations in the future.


There are various kinds of people in a society, from those who receive good reputations from many people to those who receive good reputations from a few. Such diversity is established by complex dynamics of mutual evaluation of their behavior. This study theoretically approached such complex dynamics. Although there are some differences between our simple model and a real society, our findings give some basic insight into the mechanism of how good and bad individuals emerge in the context of indirect reciprocity (corresponding to ``generalized exchange'' \cite{levi1949structures,ekeh1974social,yamagishi1993generalized,takahashi2000emergence} in sociology). In conclusion, this study provides a new theoretical approach to investigate reputation structure in the population where individuals privately assess each other.

\section*{Acknowledgments}
Y.F. acknowledges the support by JSPS KAKENHI Grant Number JP21J01393. H.O. acknowledges the support by JSPS KAKENHI Grant Number JP19H04431.

\section*{Author Contributions}
Y.F. and H.O. designed research, calculated results numerically and analytically, and wrote the paper.

\section*{Availability of Data and Materials}
The datasets used and/or analysed during the current study available from the corresponding author on reasonable request.



\newpage
\renewcommand{\theequation}{S\arabic{equation}}
\setcounter{equation}{0}
\renewcommand{\figurename}{FIG. S}
\setcounter{figure}{0}

\section*{Supplementary Information}
Here, we show a detailed calculation of triples $\{(q_j, \mu_j,\sigma_j)\}_{j=1, \cdots}$ which satisfy Eq.~(\ref{equilibrium_g2}).
\vspace{0.7cm}

{\bf When the social norm is SJ}: In the main text, we have already derived
\begin{equation}
    \mu_1=f_{\rm SJ}^{\rm C}(\mu_1)=f_{\rm SJ}^{\rm D}(\mu_1)=\frac{1}{2}.
\label{mu_SJ}
\end{equation}
It is trivial that the mass $q_1$ is
\begin{equation}
    q_1=1.
\end{equation}
We now also derive variance $\sigma_1^{2}$. When we substitute $\mu_1$ and $q_1$ above into Eq.~(\ref{equilibrium_g2}), we obtain
\begin{equation}
\begin{split}
    g(p;\mu_1,\sigma_1^2)&=h(\mu_1)g(p;f_{\rm SJ}^{\rm C}(\mu_1),s^2+(\underbrace{\Delta f_{\rm SJ}^{\rm C}}_{=1-2e_2})^2\sigma_1^2)+(1-h(\mu_1))g(p;f_{\rm SJ}^{\rm D}(\mu_1),s^2+(\underbrace{\Delta f_{\rm SJ}^{\rm D}}_{=-(1-2e_2)})^2\sigma_1^2)\\
    &=h(\mu_1)g(p;\mu_1,s^2+(1-2e_2)^2\sigma_1^2)+(1-h(\mu_1))g(p;\mu_1,s^2+(1-2e_2)^2\sigma_1^2)\\
    &=g(p;\mu_1,s^2+(1-2e_2)^2\sigma_1^2).
\end{split}
\end{equation}
By comparing terms of variances between the left and right sides of this equation, we obtain
\begin{equation}
\begin{split}
    &\sigma_1^2=s^2+(1-2e_2)^2\sigma_1^2,\\
    &\Leftrightarrow \sigma_1^2= s^{2} \frac{1}{1-(1-2e_2)^2} = \frac{e_{2}(1-e_{2})}{N}\frac{1}{1-(1-2e_2)^2} = \frac{1}{4N}.
\end{split}
\end{equation}
\vspace{0.7cm}

{\bf When the social norm is SS}: In the main text, we have already derived
\begin{equation}
\begin{split}
    &\mu_1=1-e_2,\\
    &\mu_{j+1}=f_{\rm SS}^{\rm D}(\mu_j)\ \ (j \ge 1).
\end{split}
\end{equation}
This recurrence relation can be analytically solved as
\begin{equation}\label{mu_SS}
\begin{split}
    &\mu_{j+1}=-(1-2e_2)\mu_j+(1-e_2)\\
    &\Leftrightarrow \left(\mu_{j+1}-\frac{1}{2}\right)=-(1-2e_2)\left(\mu_j-\frac{1}{2}\right)\\
    &\Leftrightarrow \mu_j=\{-(1-2e_2)\}^{j-1}\left(\mu_1-\frac{1}{2}\right)+\frac{1}{2}\\
    &\hspace{0.91cm}=\frac{1-\{-(1-2e_2)\}^j}{2}.
\end{split}
\end{equation}

Now we derive variance $\sigma_j^{2}$ and mass $q_j$. When we substitute Eq.~(\ref{mu_SS}) into Eq.~(\ref{equilibrium_g2}), we obtain
\begin{equation}
\begin{split}
    \sum_{j=1}^{\infty}q_jg(p;\mu_j,\sigma_j^2)&=\sum_{j=1}^{\infty}q_j\{h(\mu_j)g(p;f_{\rm SS}^{\rm C}(\mu_j),s^2+(\underbrace{\Delta f_{\rm SS}^{\rm C}}_{=0})^2\sigma_j^2)+(1-h(\mu_j))g(p;f_{\rm SS}^{\rm D}(\mu_j),s^2+(\underbrace{\Delta f_{\rm SS}^{\rm D}}_{=-(1-2e_2)})^2\sigma_j^2)\}\\
    &=\sum_{j=1}^{\infty}q_j\{h(\mu_j)g(p;\mu_1,s^2)+(1-h(\mu_j))g(p;\mu_{j+1},s^2+(1-2e_2)^2\sigma_j^2)\}\\
    &=\left(\sum_{j=1}^{\infty}q_jh(\mu_j)\right)g(p;\mu_1,s^2)\sum_{j=1}^{\infty}q_j(1-h(\mu_j))g(p;\mu_{j+1},s^2+(1-2e_2)^2\sigma_j^2).
\label{equilibrium_g_SS}
\end{split}
\end{equation}
By comparing terms between the left and right sides of Eq.~(\ref{equilibrium_g_SS}), we obtain a recurrence relation for the variances $\sigma_j^2$ as
\begin{equation}\label{sigma_recurrence_SS}
\begin{split}
    &\sigma_1^2=s^2,\\
    &\sigma_{j+1}^2=s^2+(1-2e_2)^2\sigma_j^2\ \ (j\ge 1).
\end{split}
\end{equation}
This recurrence relation can be solved as
\begin{equation}
\begin{split}
    &\sigma_{j+1}^2=s^2+(1-2e_2)^2\sigma_j^2\\
    &\Leftrightarrow \left(\sigma_{j+1}^2-\frac{s^2}{1-(1-2e_2)^2}\right)=(1-2e_2)^2\left(\sigma_j^2-\frac{s^2}{1-(1-2e_2)^2}\right)\\
    &\Leftrightarrow \sigma_j^2=(1-2e_2)^{2(j-1)}\left(\sigma_1^2-\frac{s^2}{1-(1-2e_2)^2}\right)+\frac{s^2}{1-(1-2e_2)^2}\\
    &\hspace{0.95cm}=s^2\frac{1-(1-2e_2)^{2j}}{1-(1-2e_2)^2}=\frac{e_{2}(1-e_{2})}{N}\frac{1-(1-2e_2)^{2j}}{1-(1-2e_2)^2}=\frac{1-(1-2e_2)^{2j}}{4N}.
\end{split}
\end{equation}
Similarly, by comparing terms between the left and right sides of Eq.~(\ref{equilibrium_g_SS}), we can obtain a recurrence relation for the masses $q_j$ as
\begin{equation}
\begin{split}
    &\left\{\begin{array}{l}
        q_1=\sum_{j=1}^{\infty} q_jh(\mu_j)\\
        q_{j+1}=q_j(1-h(\mu_j)).\\
    \end{array}\right.\\
\end{split}
\end{equation}
By using $\sum_{j=1}^{\infty} q_{j}=1$, this is solved as
\begin{align}
    & q_j=\frac{\prod_{k=1}^{j-1}(1-h(\mu_k))}{\sum_{\ell=1}^{\infty} \prod_{k=1}^{\ell-1}(1-h(\mu_k))},
\end{align}
where and hereafter we use the convention, $\prod_{k=1}^{0} \cdot = 1$.
\vspace{0.7cm}

{\bf When the social norm is SH}: In the main text, we have already derived
\begin{equation}
\begin{split}
    &\mu_1=e_2,\\
    &\mu_{j+1}=f_{\rm SH}^{\rm C}(\mu_j)\ \ (j \ge 1).
\end{split}
\end{equation}
This recurrence relation can be analytically solved as
\begin{equation}\label{mu_SH}
\begin{split}
    &\mu_{j+1}=(1-2e_2)\mu_j+e_2\\
    &\Leftrightarrow \left(\mu_{j+1}-\frac{1}{2}\right)=(1-2e_2)\left(\mu_j-\frac{1}{2}\right)\\
    &\Leftrightarrow \mu_j=(1-2e_2)^{j-1}\left(\mu_1-\frac{1}{2}\right)+\frac{1}{2}\\
    &=\frac{1-(1-2e_2)^j}{2}.
\end{split}
\end{equation}

We also derive variance $\sigma_j^2$ and mass $q_j$. When we substitute Eq.~(\ref{mu_SH}) into Eq.~(\ref{equilibrium_g2}), we obtain
\begin{equation}
\begin{split}
    \sum_{j=1}^{\infty}q_jg(p;\mu_j,\sigma_j^2)&=\sum_{j=1}^{\infty}q_j\{h(\mu_j)g(p;f_{\rm SH}^{\rm C}(\mu_j),s^2+(\underbrace{\Delta f_{\rm SH}^{\rm C}}_{=1-2e_2})^2\sigma_j^2)+(1-h(\mu_j))g(p;f_{\rm SH}^{\rm D}(\mu_j),s^2+(\underbrace{\Delta f_{\rm SH}^{\rm D}}_{=0})^2\sigma_j^2)\}\\
    &=\sum_{j=1}^{\infty}q_j\{h(\mu_j)g(p;\mu_{j+1},s^2+(1-2e_2)^2\sigma_j^2)+(1-h(\mu_j))g(p;\mu_1,s^2)\}\\
    &=\sum_{j=1}^{\infty}q_jh(\mu_j)g(p;\mu_{j+1},s^2+(1-2e_2)^2\sigma_j^2)+\left(\sum_{j=1}^{\infty}q_j(1-h(\mu_j))\right)g(p;\mu_1,s^2).
\label{equilibrium_g_SH}
\end{split}
\end{equation}
By comparing terms between the left and right sides of Eq.~(\ref{equilibrium_g_SH}), we obtain a recurrence relation for the variances $\sigma_j^2$ as
\begin{equation}
\begin{split}
    &\sigma_1^2=s^2,\\
    &\sigma_{j+1}^2=s^2+(1-2e_2)^2\sigma_j^2\ \ (j\ge 1).
\end{split}
\end{equation}
Because this recurrence relation is same as Eq.~(\ref{sigma_recurrence_SS}) for the case of $A={\rm SS}$, we obtain
\begin{equation}
    \sigma_j^2=\frac{1-(1-2e_2)^{2j}}{4N}.
\end{equation}
Similarly, by comparing terms between the left and right sides of Eq.~(\ref{equilibrium_g_SH}), we can obtain a recurrence relation for the masses $q_j$ as
\begin{equation}
\begin{split}
    &\left\{\begin{array}{l}
        q_1=\sum_{j=1}^{\infty} q_j(1-h(\mu_j))\\
        q_{j+1}=q_jh(\mu_j).\\
    \end{array}\right.
\end{split}
\end{equation}
By using $\sum_{j=1}^{\infty} q_{j}=1$, this is solved as
\begin{align}
   q_j=\frac{\prod_{k=1}^{j-1}h(\mu_k)}{\sum_{\ell=1}^{\infty} \prod_{k=1}^{\ell-1}h(\mu_k)}. 
\end{align}
\vspace{0.7cm}

{\bf When the social norm is SC}: In the main text, we have already derived
\begin{equation}\label{mu_SC}
\begin{split}
    &\mu_1=1-e_2,\\
    &\mu_2=e_2.\\
\end{split}
\end{equation}

We also derive variance $\sigma_j^{2}$ and mass $q_j$. When we substitute Eq.~(\ref{mu_SC})  into Eq.~(\ref{equilibrium_g2}), we obtain
\begin{equation}
\begin{split}
    \sum_{j=1}^{2}q_jg(p;\mu_j,\sigma_j^2)&=\sum_{j=1}^{2}q_j\{h(\mu_j)g(p;f_{\rm SC}^{\rm C}(\mu_j),s^2+(\underbrace{\Delta f_{\rm SC}^{\rm C}}_{=0})^2\sigma_j^2)+(1-h(\mu_j))g(p;f_{\rm SC}^{\rm D}(\mu_j),s^2+(\underbrace{\Delta f_{\rm SC}^{\rm D}}_{=0})^2\sigma_j^2)\}\\
    &=\sum_{j=1}^{2}q_j\{h(\mu_j)g(p;\mu_1,s^2)+(1-h(\mu_j))g(p;\mu_2,s^2)\}\\
    &=\left(\sum_{j=1}^{2}q_jh(\mu_j)\right)g(p;\mu_1,s^2)+\left(\sum_{j=1}^{2}q_j(1-h(\mu_j))\right)g(p;\mu_2,s^2),
\label{equilibrium_g_SC}
\end{split}
\end{equation}
By comparing terms between the left and right sides of Eq.~(\ref{equilibrium_g_SC}), we obtain the variances $\sigma_j^2$ as
\begin{equation}
    \sigma_1^2=\sigma_2^2=s^2=\frac{e_{2}(1-e_{2})}{N}
\end{equation}
Similarly, by comparing terms between the left and right sides of Eq.~(\ref{equilibrium_g_SC}), we obtain the relation that the masses $q_1$ and $q_{2}$ satisfy, as
\begin{equation}
\begin{split}
    &\left\{\begin{array}{l}
        q_1=q_1h(\mu_1)+q_2h(\mu_2)\\
        q_2=q_1(1-h(\mu_1))+q_2(1-h(\mu_2)).\\
    \end{array}\right.
\end{split}
\end{equation}
By using $q_{1}+q_{2}=1$, this is solved as
\begin{align}
    &q_1=q_2=\frac{1}{2}.
\end{align}

\end{document}